\documentclass[oldversion]{aa}     
\def \th {\thinspace}

\def\approxgt{\mathrel{\hbox{\rlap{\lower.55ex \hbox {$\sim$}} \kern-.3em \raise.4ex \hbox{$>$}}}}
\def\lesssim{\mathrel{\hbox{\rlap{\lower.55ex \hbox {$\sim$}} \kern-.3em \raise.4ex \hbox{$<$}}}}
\def\approxlt{\mathrel{\hbox{\rlap{\lower.55ex \hbox {$\sim$}} \kern-.3em \raise.4ex \hbox{$<$}}}}
\def \degmark {^\circ}

\usepackage{epsf}
\usepackage{graphicx}
\begin{document}

\title{On the nature of the Cygnus\th X-2 like Z-track sources}

\author{M. Ba\l uci\'nska-Church\inst{1,2}
\and A. Gibiec\inst{2}
\and N. K. Jackson\inst{1}
\and M. J. Church\inst{1,2}}
\institute{School of Physics and Astronomy, University of Birmingham,
           Birmingham, B15 2TT, UK\\
\and
          Astronomical Observatory, Jagiellonian University,
          ul. Orla 171, 30-244 Cracow, Poland.\\}
\offprints{mbc@star.sr.bham.ac.uk}

\date{Received August 2009; Accepted January 8 2010}
\titlerunning{The nature of the Cyg-like sources}
\authorrunning{ Ba\l uci\'nska-Church et al.}

\abstract{
Based on the results of applying the extended ADC emission model for low mass X-ray binaries 
to three Z-track sources: GX\th 340+0, GX\th 5-1 and \hbox{Cyg\th X-2}, we propose an explanation 
of the Cygnus\th X-2 like Z-track sources. The Normal Branch is dominated by the increasing radiation 
pressure of the neutron star caused by a mass accretion rate that increases between the soft apex and 
the hard apex. The radiation pressure continues to increase on the Horizontal Branch becoming several 
times super-Eddington. We suggest that this disrupts the inner accretion disk and that part of the 
accretion flow is diverted vertically forming jets which are detected by their radio emission on this 
part of the Z-track. We thus propose that high radiation pressure is the necessary condition for the 
launching of jets. On the Flaring Branch there is a large increase in the neutron star blackbody 
luminosity at constant mass accretion rate indicating an additional energy source on the neutron star. 
We find that there is good agreement between the mass accretion rate per unit emitting area of the 
neutron star $\dot m$ at the onset of flaring and the theoretical critical value at which burning becomes 
unstable. We thus propose that flaring in the Cygnus\th X-2 like sources consists of unstable nuclear 
burning. Correlation of 
measurements of kilohertz QPO frequencies in all three sources with spectral fitting results leads to 
the proposal that the upper kHz QPO is an oscillation always taking place at the inner accretion disk 
edge, the radius of which increases due to disruption of the disk by the high radiation pressure of the 
neutron star.
\keywords{Accretion: accretion disks -- acceleration of particles -- binaries:
close -- stars: neutron -- X-rays: binaries -- X-rays: individual (Cyg\th X-2, GX\th 340+0, GX\th 5-1)}}
\maketitle

%
\section{Introduction}

The Z-track sources are the brightest group of Galactic low mass X-ray binaries (LMXB) containing 
a neutron star persistently emitting at the Eddington luminosity or several times this. The sources 
trace out a Z-shape in hardness-intensity or hardness-softness (Hasinger \& van der Klis 1989) 
having three branches labelled the horizontal branch (HB), the normal branch (NB) and flaring 
branch (FB). Strong physical changes are clearly taking place at the inner disk and neutron star, 
but a generally accepted explanation of the Z-track phenomenon has not existed. Hasinger \& 
van der Klis (1989) presented colour-colour diagrams for all six original Z-track sources using 
{\it Exosat} data which showed rather different shaped Z-tracks between two groups: Cyg\th X-2, 
GX\th 340+0 and GX\th 5-1 (the Cygnus\th X-2 like sources) and Sco\th X-1, GX\th 17+2 and GX\th 349+2 
(the Sco\th X-1 like sources). It was proposed that these constituted two sub-groups of Z-track 
sources, for example by Kuulkers et al. (1994), and this became generally accepted (Kuulkers \&
van der Klis 1995; Smale et al. 2000, 2003). Many observations have confirmed the rather different 
shaped Z-tracks, the Cygnus\th X-2 like sources having a long HB, while the Sco\th X-1 like sources
have a short HB but a long FB.

LMXB have been classified according to the various phenomena observed, into the Atoll, Z-track, dipping
and ADC sources, and this has been shown to relate to inclination angle so that X-ray dipping is 
seen at inclinations greater than $\sim$65$\degmark$. The Z-track sources display
characteristic patterns in hardness-intensity but, in general, no orbital related behaviour indicating smaller 
inclination and the Atoll sources exhibit rather different patterns of hardness-intensity.
However, a more basic classification would be based on luminosity in which the Z-track sources
have luminosities greater than $\sim$10$^{38}$ erg s$^{-1}$ and the Atoll sources have luminosities
between $\sim$10$^{36}$ and $\sim$10$^{38}$ erg s$^{-1}$, although not all lower luminosity LMXB 
have been identified as Atoll sources via their hardness-intensity patterns. These patterns are also 
not understood, nor is the relation between the Z-track and Atoll classes, so that understanding 
Z-track and Atoll sources remains a fundamental problem.

Thus we need to understand the detailed nature of the inner disk and accretion in the super-Eddington 
LMXB, the Z-track sources, but these are also important since they are detected as radio emitters, 
although essentially in one part of
the Z-track only: the horizontal branch. Not only is radio detected, but striking results from the VLA show the 
release of massive radio condensations from the source Sco\th X-1 (Fomalont et al. 20001) moving 
outwards in two opposite direction at velocity v/c $\sim$ 0.45. Because radio is detected essentially 
in one branch only, the sources offer the possibility of determining conditions pertaining inside the sources
via X-ray observations, i.e. the conditions at the inner disk, when jets are present, which distinguish 
the horizontal branch from the other branches, and so finding the conditions necessary for jet formation.  

Possible ways of understanding the Z-track sources are by timing studies, spectral studies or 
theoretical studies. Extensive timing studies have been made to investigate the quasi periodic
oscillations (QPO) found in the Z-track and other sources (e.g. van der Klis et al. 1987). This 
has revealed a wealth of complex behaviour, and in particular, study of the kilohertz QPO has revealed
a pattern of variation of the twin kHz peaks on the horizontal branch of Z-track sources, and
various models have been proposed to explain the QPO (Stella \& Vietri 1998; Abramowicz \&
Klu\'zniak 2001; Miller et al. 1998). However, it is difficult to determine what changes take 
place at the inner disk around the Z-track from data on the kHz QPO. 

Previous spectral fitting tended to apply a particular LMXB emission model: the Eastern model 
of Mitsuda et al. (1989) in which the emission is assumed to consist of disk blackbody emission
plus Comptonized emission from the neutron star or inner disk. Application of this model may 
have given acceptable fits to the spectra (Done et al. 2002, Agrawal \& Sreekumar 2003; 
di Salvo et al. 2002), however, interpretation of the results, i.e. of the variation of the 
various spectral parameters around the Z-track was not easy. Our previous work over an 
extended period on the dipping class of LMXB provided evidence for a different emission model 
and we proposed the ``extended ADC (accretion disk corona)'' model 
(Church \& Ba\l uci\'nska-Church 2004, 1995) in 
which the emission components consist of blackbody emission from the neutron star plus 
Comptonized emission from an extended ADC. In this model, the ADC covers a substantial
fraction of the inner disk and the thermal emission of the disk provides the seed photons
for Comptonization (Church \& Ba\l uci\'nska-Church 2004).
The model was developed from longterm studies of the dipping LMXB sources. 
In all LMXB, the observed shape of the spectrum and spectral fitting 
show that Comptonized emission dominates the spectrum, i.e. comprises more than 90\% of
the total luminosity, as shown for example in the {\it ASCA} survey of LMXB of Church 
\& Ba\l uci\'nska-Church (2001). In less luminous sources, the percentage may be higher, while in
the brightest sources, the Z-track sources, the contribution is $\sim$90\% at the soft apex. 
In the dipping sources, it is generally accepted that the reductions in X-ray 
intensity at the orbital period are due to absorption in the bulge in the outer accretion disk 
(White \& Swank 1982; Walter et al. 1982). Spectral fitting of non-dip and dip spectra clearly 
shows that the dominant Comptonized spectral component is removed only slowly in dipping and 
so the 90\% contributor to the total luminosity must be extended (e.g. Church et al. 1997).
Measurement of the dip ingress times provides the sizes of the extended emission. These measurements 
indicated the radial extent to be very large: between 20\th 000 and 700\th 000 km depending on source 
luminosity (Church \& Ba\l uci\'nska-Church 2004). Moreover, this extended region cannot reach very
high above the disk otherwise it would not be possible to fully cover the emitter by absorber
and 100\% deep dipping, often observed, would not be possible. Thus the conclusion cannot be avoided that
90\% of the total luminosity originates in this extended region which, given the measured geometry:
large in radial extent but small in height, can be identified with a hot, thin 
accretion disk corona above the disk, typically extending across 15\% of the disk radius.
Recently independent support for 
the model came from the {\it Chandra} grating results of Schulz et al. (2009) on Cyg\th X-2. Precise 
grating measurements revealed a wealth of emission lines of highly excited states such as the 
H-like ions of Ne, Mg, Si, S and Fe. The width of these lines indicated Doppler shifts due to 
orbital motion in the accretion disk corona at radial positions between 18\th 000 and 110\th 000 km 
in good agreement with the overall ADC size from dip ingress timing. Further support for line formation
at large radial distances in an extended ADC came from recent results of {\it Chandra} observations of 
4U\th 1624-490 (Xiang et al. 2009). The evidence for the extended 
corona is thus now overwhelming.

Theoretical modelling for the Z-track sources was carried out by Psaltis et al. (1995) assuming a
magnetosphere of the neutron star and the changing properties and geometry of this as the
mass accretion rate changed. However, the model assumed that the Comptonized emission observed 
in the spectra of all LMXB originated in a small central region close to the neutron star,
not in an extended region.

It should also be stated here that although 
the nature of the Z-track has not been clear, there has been a standard
assumption made that the mass accretion rate increases monotonically around the Z-track
in the sense HB - NB - FB, based originally on results from multi wavelength observations
of Cygnus\th X-2 (Hasinger et al. 1990). This assumption has been widely but not universally
adopted and it will appear in Sects. 3 and 4 that our results do not support this assumption.

More recently, we have taken the approach of applying the extended ADC model to the Z-track sources
GX\th 340+0 (Church et al. 2006) and GX\th 5-1 (Jackson et al. 2009), on the basis that the
evidence favours the Extended ADC model, to test the hypothesis that this can fit the spectra,
and provide a physically reasonable and consistent explanation of the Z-track.
Good fits to the spectra at all 
positions on the Z-track were obtained, and the spectral fitting clearly suggested an explanation of 
the physical changes taking place on the Z-track. At the Soft Apex: the junction of the HB and FB, 
the X-ray intensity is minimum. Ascending the normal branch to the Hard Apex, the intensity increases 
substantially and the neutron star blackbody temperature $kT$ increases from $\sim$1 to $\sim$2 keV. 
The luminosity of the ADC Comptonized emission ($L_{\rm ADC}$) which is  90\% of the total 1 - 30 keV 
luminosity at 
the soft apex increases by a factor of two or more strongly suggesting that $\dot M$ was increasing. 
The increase of $kT$ is consistent with this and has a major consequence that the radiation pressure
of the neutron star ($\sim T^4$) increases by an order of magnitude. The strength of the radiation
pressure close to the neutron star which will determine its effects on the inner accretion disk is
conveniently expressed in terms of the ratio of the flux emitted per unit area to its Eddington value 
$f/f_{\rm Edd}$ where $f_{\rm Edd}$ = $L_{\rm Edd}/4\,\pi \, R^2$ and $R$ is the radius of the neutron 
star. This ratio is a better measure of radiation pressure effects close to the neutron star than 
$L/L_{\rm Edd}$ which averages the flux over the surface, especially if only part of the surface is 
emitting. In GX\th 340+0, $f/f_{\rm Edd}$ was found to increase from 0.2 at the soft apex to 1.5
at the hard apex rising to almost 3 times super-Eddington on the HB. In GX\th 5-1 the ratio similarly
increased to become greater than unity, so that in both sources, the radiation became super-Eddington
at the hard apex and HB. It is these parts of the Z-track on which jets are observed
and we proposed that the high radiation pressure disrupts the 
\begin{figure*}[!ht]                                                           
\begin{center}
\includegraphics[width=60mm,height=140mm,angle=270]{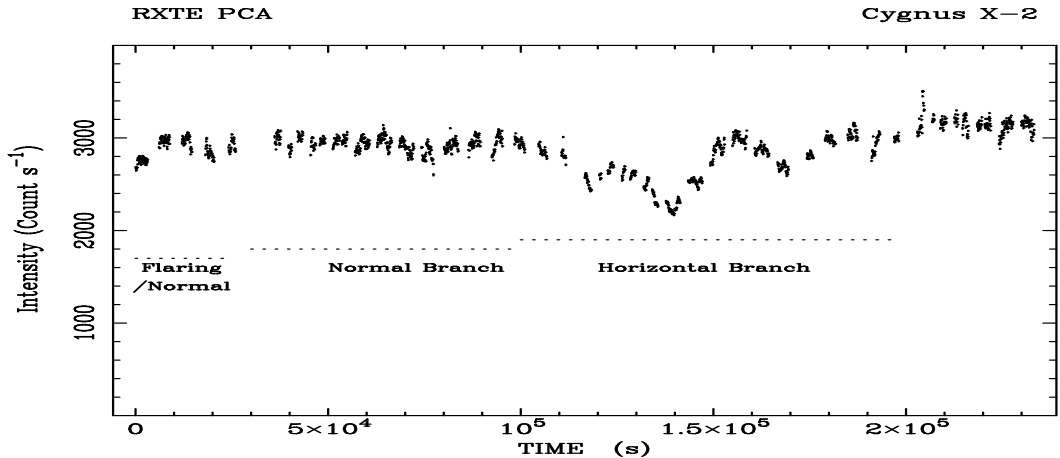}                  
\caption{Background-subtracted and deadtime-corrected PCA light curve of
Cygnus\th X-2 with 64 s binning.}
\label{}
\end{center}
\end{figure*}
inner disk and drives matter
into the vertical direction so launching the jets (Church et al. 2006; Jackson et al. 2009).

In both sources, on the flaring branch $L_{\rm ADC}$ was constant suggesting $\dot M$ was constant, 
while the neutron star blackbody luminosity increased strongly implying that a non-accretion
powered energy source developed on the neutron star. The mass accretion rate per unit area 
of the neutron star $\dot m$ =  $\dot M/4\, \pi\, R_{\rm BB}^2$, where $R_{\rm BB}$ is the blackbody 
radius derived from spectral fitting, was found at the start of the FB, i.e. the soft apex, 
to be in good agreement with the theoretical critical value $\dot m_{\rm ST}$ (Bildsten 1998).
This is the critical value of $\dot m$ below which nuclear burning on the surface of the neutron star
becomes unstable, i.e. for $\dot m$ $<$ $1.3\times 10^5$ g cm$^{-2}$ s$^{-1}$ the burning changes
from stable to unstable. Thus it was proposed that for a source descending the normal branch
with falling $\dot M$, the onset of unstable burning causes the strong flaring
seen in the X-ray lightcurve. The increase in blackbody luminosity is simply the nuclear
burning power, as seen at essentially constant $L_{\rm ADC}$, i.e. constant $\dot M$.
 
In the second source, GX\th 5-1, the evolution of the kilohertz QPO was also investigated
by Jackson et al. (2009). Using the same selections of data used in spectral analysis, the variation
of the kHz QPO along the HB and part of the NB was obtained by timing analysis, and viewed in
the light of the spectral fitting results.
It is well-known that the upper and lower kHz QPO frequencies $\nu_2$ and $\nu_1$
vary on the HB (e.g. Jonker et al. 2000; Wijnands et al. 1998a,b), 
and that if $\nu_2$ is assumed to correspond to a Keplerian
frequency in the disk, the radial position $r$ is $\sim$18 km, i.e. some way into the disk. A 
previously unknown
correlation was found between the measured frequencies and the above ratio $f/f_{\rm Edd}$ 
(Jackson et al. 2009), specifically that the $r$ was approximately constant until $f/f_{\rm Edd}$
= 1, above which $r$ increased rapidly. This supports the proposal of disk disruption by high
radiation pressure; moreover it suggests a mechanism for the upper frequency kHz QPO that this 
oscillation is always an oscillation at the inner disk edge, the radius of which was modified by 
different levels of radiation pressure. Thus disruption of the inner disk was the factor determining 
the frequency of the higher frequency kHz QPO and its variation.

In this work, we apply the extended ADC model for the first time to the Z-track source 
Cygnus\th X-2, also determining the evolution of the kHz QPO around the Z-track in Cyg\th X-2
and in GX\th 340+0 and we present a comparison of the three Cygnus\th X-2 like sources.

\section{Observations and analysis}

We analysed the {\it Rossi-XTE} observation of Cygnus\th X-2 of June 30 - July 3, 1997
in which a full Z-track was covered in a relatively short time. The observation consisted
of 11 sub-observations spanning 235 ksec. Data from both the proportional counter array 
(PCA: 2 - 60 keV) and the high energy X-ray timing experiment (HEXTE: 15 - 250 keV) were 
used. PCA data in Standard2 mode with 16 s resolution were used for the extraction of light
curves and spectra. Examination of the housekeeping data showed that three of the five Xe 
proportional counter units (PCUs) were consistently on, and so data were extracted from these 
(PCUs 0, 1 and 2). Standard screening criteria were applied to select data with an offset
between the source and telescope pointing direction of less than 0.02$\degmark$ and elevation
above the Earth's limb greater than 10$\degmark$. Data were extracted from the top layer 
of the detectors using both left and right anodes. Analysis was carried out using the 
standard {\it RXTE} software {\sc ftools 5.3.1}. A lightcurve was generated in the band
1.9 -- 20.3 keV and the facility {\sc pcabackest} used to generate background files for
each raw data file based on the latest bright-source background model for Epoch 3 of the 
mission during which the observations were made. Deadtime correction was carried out, and
the background-subtracted, deadtime-corrected lightcurve is shown in Fig. 1 with 64 s
binning. Lightcurves were also extracted in three sub-bands: 1.9 - 4.1, 4.1 - 6.9
and 6.9 - 20.3 keV. By dividing the high band by the medium band lightcurve, a 
hardness-intensity plot was made as shown in Fig. 2. Individual subsections of the lightcurve were
identified with their positions on the Z-track by making a hardness-intensity  plot for
each and locating its position on the overall Z-track. Thus the three Z-track branches are 
identified as shown in Fig. 1.

Lightcurves and spectra were also extracted from Cluster 1 and Cluster 2 of the HEXTE instrument, so that
simultaneous spectral fitting of PCA and HEXTE spectra could be carried out. These were
extracted with the {\sc ftool} {\sc hxtlcurv} which also provided background files and 
allowed deadtime correction.

Timing analysis was carried out as a function of position along the Z-track. The
data used for this were single bit data of type SB\_125us\_14\_249\_1s, i.e. data having
125 $\mu$s resolution and a single energy bin spanning 5 - 60 keV. Power density spectra were
extracted for each selection on the Z-track
using the {\it Xronos} facility {\sc powspec} in a single energy bin, and Leahy 
normalization was applied. To allow high quality fitting of the spectra, they were converted 
to a format allowing use of the {\it Xspec} spectral fitting software, by modifying the file
headers and re-naming of columns.
{\sc flx2xsp} was used to produce a file with the required {\it pha} extension and to
generate a dummy instrument response as needed by {\it Xspec} consisting of a unitary matrix
having no effect on the data. The spectra were grouped to a minimum count per bin to allow use 
of the $\chi^2$ statistic, the grouping varying between 40 and 140 counts per bin
as found appropriate by trial and error.

Timing analysis was also carried out for the source GX\th 340+0 (Sect. 5) so that in all three sources
timing analysis results could be correlated with spectral fitting results for the same
selections of data along the Z-track.
In the case of GX\th 340+0, the data used were event mode data of type E\_8us\_32B\_14\_1s
which were fitted in the range 5 - 60 keV.
The data were regrouped appropriately, again having 4 - 140 counts per bin.
 
Following presentation of results for Cygnus\th X-2, we present a comparison of
spectral fitting and of timing analysis results for all
three Cygnus\th X-2 like sources: \hbox{Cyg\th X-2,} GX\th 340+0 and GX\th 5-1 using our
previous analyses of the {\it RXTE} observations of GX\th 340+0 made on 1997 September lasting 400
ksec and the 1998 November observation of GX\th 5-1 spanning 95 ksec (Church et al. 2006; 
Jackson et al. 2009). Results for the 1997 June observation of GX\th 340+0 are also shown. 

\section{Results: Cygnus\th X-2}

The PCA lightcurve shown in Fig. 1 shows that in the first 25 ksec of the observation, the source
was in a period of flaring, moving on the Z-track along the flaring branch, and then onto the
normal branch. The source stayed on the normal branch, until at $\sim$100 ksec
\begin{figure}[!ht]                                                           
\begin{center}
\includegraphics[width=74mm,height=74mm,angle=270]{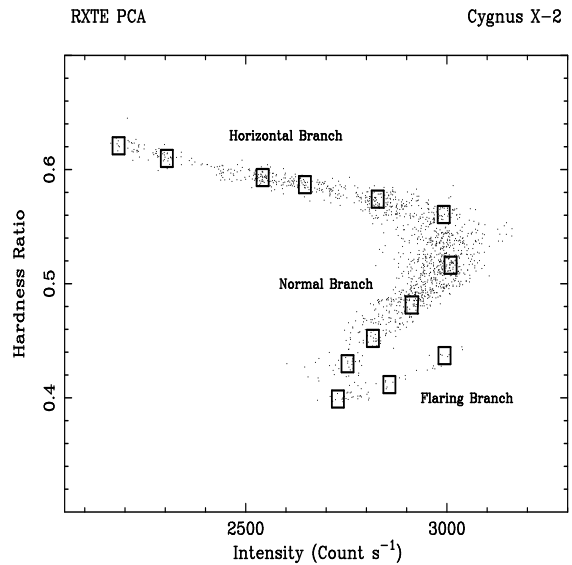}     
\caption{Hardness ratio \hbox{(7.3 -- 18.1 keV)/(4.1 -- 7.3 keV)} variation with intensity.}
\label{}
\end{center}
\end{figure}
it moved to the horizontal branch. It can be seen that at about 200 ksec 
\tabcolsep 3.5 mm
\begin{table*}
\begin{center}
\caption{Spectral fitting results for Cygnus\th X-2: 90\% confidence errors are shown.}
\begin{minipage}{160mm}
\begin{tabular}{lrrrrlrrr}
\hline \hline
$\;\;$spectrum&$N_{\rm H}$&$kT$&$norm_{\rm BB}$&$R_{\rm BB}$&$E_{\rm
CO}$&$norm_{\rm CPL}$&$\chi^2$/d.o.f.\\
&&(keV)&&(km)&(keV)\\
\noalign{\smallskip\hrule\smallskip}
Horizontal Branch\\
1&1.5$\pm 0.6$  &2.31$\pm 0.20$  &1.96$\pm 0.33$ &2.1$\pm 0.4$ &7.29$\pm 0.69$  &3.93$\pm 0.41$  &66.7/69\\
2&2.0$\pm 0.5$  &2.37$\pm 0.16$ &1.74$\pm 0.24$ &1.9$\pm 0.3$ &7.00$\pm 0.45$  &4.50$\pm 0.34$  &77.6/68\\
3&2.2$\pm 0.5$  &2.20$\pm 0.14$  &1.67$\pm 0.20$ &2.1$\pm 0.3$  &6.90$\pm 0.27$  &5.16$\pm 0.33$  &64.3/68\\
4&2.3$\pm 0.5$  &2.20$\pm 0.15$  &1.76$\pm 0.20$ &2.2$\pm 0.3$  &6.70$\pm 0.30$  &5.52$\pm 0.39$  &82.8/68\\
5&2.6$\pm 0.5$  &1.99$\pm 0.16$  &1.51$\pm 0.30$ &2.4$\pm 0.5$  &6.68$\pm 0.24$  &6.15$\pm 0.45$  &73.4/65\\
\noalign{\smallskip}
Hard Apex\\
6&2.7$\pm 0.5$  &1.92$\pm 0.15$  &1.60$\pm 0.35$ &2.7$\pm 0.5$  &6.43$\pm 0.22$  &6.71$\pm 0.50$  &52.6/55\\
\noalign{\smallskip}
Normal Branch \\
7&2.6$\pm 0.5$  &1.65$\pm 0.12$  &1.43$\pm 0.50$ &3.5$\pm 0.7$   &5.95$\pm 0.16$  &7.28$\pm 0.60$  &78.4/68\\
8&2.4$\pm 0.5$  &1.46$\pm 0.08$  &1.56$\pm 0.60$ &4.7$\pm 1.0$   &5.68$\pm 0.18$  &7.23$\pm 0.72$  &56.8/58\\
9   &1.9$\pm 0.6$  &1.26$\pm 0.05$  &2.54$\pm 0.75$ &7.9 $\pm 1.3$ &5.59$\pm 0.26$   &6.35$\pm 0.85$ &60.3/51 \\
10&1.6$\pm 0.6$ &1.20$\pm 0.04$  &2.93$\pm 0.75$ &9.5$\pm 1.3$   &5.45$\pm 0.30$  &6.1$\pm 0.9$  &74.1/70\\
\noalign{\smallskip}
Soft Apex \\
11&0.9$\pm 0.6$  &1.19$\pm 0.03$  &4.55$\pm 0.70$ &12.0$\pm 1.1$ &5.41$\pm 0.30$  &4.97$\pm 0.85$  &53.6/63\\
\noalign{\smallskip}
Flaring Branch\\
12&0.9$\pm 0.9$ &1.22$\pm 0.04$ &4.80$\pm 1.30$ &11.9$\pm 1.0$ &5.53$\pm 0.60$  &5.05$\pm 1.50$  &47.9/47 \\
13&0.9$\pm 0.9$ &1.31$\pm 0.03$ &5.46$\pm 0.6$ &10.9$\pm 0.8$ &5.72$\pm 0.26$  &4.87$\pm 0.50$  &58.5/56\\
\noalign{\smallskip}\hline
\end{tabular}\\
Column densities are in units of 10$^{22}$ atom cm$^{-2}$; the normalization of the blackbody is in units
of $\rm {10^{37}}$ erg s$^{-1}$ for a distance of 10 kpc, the cut-off power law normalization is
in units of photon cm$^{-2}$ s$^{-1}$ keV$^{-1}$ at 1 keV.
\end{minipage}
\end{center}
\end{table*}
the intensity
of the source had drifted upwards in comparison with the previous parts of the observation.
Plotting these data on a hardness-intensity diagram revealed that the whole Z-track
was moving sideways forming a ``parallel track'' which is often seen. For this reason, we
excluded these data from our analysis to avoid superimposing data in which the source has changed.
Similarly, we excluded these data from the Z-track shown in Fig. 2. A spike in the data may
also be seen at $\sim$210 ksec lasting about \hbox{600 s,} the nature of which is not clear,
however, this event occurred in the data excluded from analysis.

Spectra were extracted at 13 positions about equally spaced along the Z-track allowing for data
in some parts of the horizontal branch being sparse. Narrow ranges of hardness ratio 
(0.015 wide) and intensity (30 count s$^{-1}$ wide) were defined and incorporated in good time 
interval (GTI) files for each selection and used to extract PCA spectra and the corresponding 
background files. Deadtime correction was made using
a local facility {\sc pcadead}. Systematic errors of 1\% were added to each channel; it was not necessary
to regroup to a minimum count in each bin to allow use of the $\chi ^2$ statistic as the count in
primitive bins was already high. A response function was generated using {\sc pcarsp}.

HEXTE spectra were extracted for each selection by using the same GTI files as for the PCA. 
The auxiliary response file (arf) of May 2000 and the response matrix file (rmf) of March 1997 
were used and the rmf was rebinned to match the actual number of channels in HEXTE spectra using the
{\sc ftools} {\sc rddescr} and {\sc rbnrmf}. The maximum energies used in the PCA and HEXTE were set
at the energies at which the source flux became less than the background flux, typically 30 keV in the
PCA and 40 keV in HEXTE.

The extended ADC emission model was applied in the spectral fitting in the form {\sc ab} $\ast${\sc (cpl + bb)},
where {\sc ab} is absorption, {\sc cpl} is a cut-off power law for Comptonization in an extended corona 
(Church \& Ba\l uci\'nska-Church 2004), and {\sc bb} is the blackbody emission of the neutron star.
It is well-known that the Comptonization cut-off energy in the bright Z-track sources is relatively
low at a few keV compared with higher values in less luminous sources. Thus there is a limited energy
range for determination of the Comptonization power law index $\Gamma$ and we have taken the approach of
fixing $\Gamma$ at 1.7, a physically reasonable value (Shapiro et al. 1976), as adopted previously
for GX\th 340+0 and GX\th 5-1 (Church et al. 2006; Jackson et al. 2009). When a best-fit was obtained, 
$\Gamma$ was freed, but the value remained close to 1.7. In addition, a Fe line was
detected at $\sim$6.7 keV with an equivalent width of about 70 eV which showed relatively little variation
along the Z-track. Good fits were obtained to all spectra, and spectral fitting results are shown in 
Table 1 and in Figs. 3 and 4.

\begin{figure}[!h]                                                               
\begin{center}
\includegraphics[width=74mm,height=74mm,angle=270]{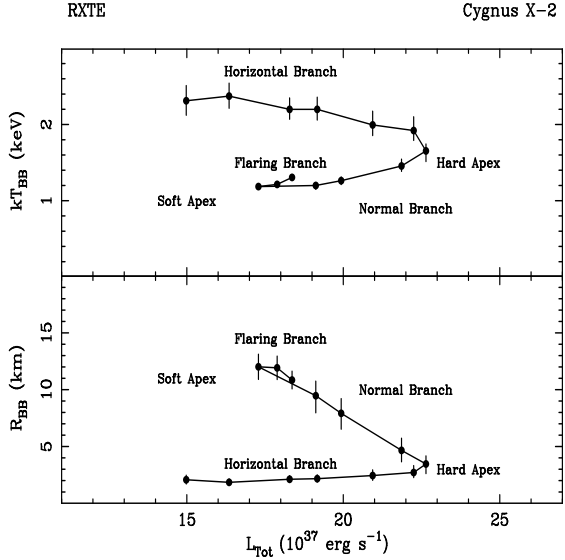}                     
\caption{Left: blackbody temperature (upper panel) and radius (lower panel) as a function of the
total luminosity.}
\end{center}
\end{figure}

In Fig. 3 we show results for the neutron star blackbody temperature $kT_{\rm BB}$ (upper panel)
and the blackbody  radius $R_{\rm BB}$ (lower panel) as functions of the total 1 - 30 keV luminosity.
A clear pattern of systematic variation is evident: the temperature is lowest at $\sim$1.3 keV at
the soft apex, suggesting that the mass accretion rate $\dot M$ is low.
At this position $R_{\rm BB}$ is maximum at 12 km,
consistent with the whole neutron star surface being emitting. On the normal branch, the temperature
rises while the blackbody radius decreases, and the temperature continues to rise on the horizontal branch.

\begin{figure}[!t]                                                               
\begin{center}
\includegraphics[width=74mm,height=74mm,angle=270]{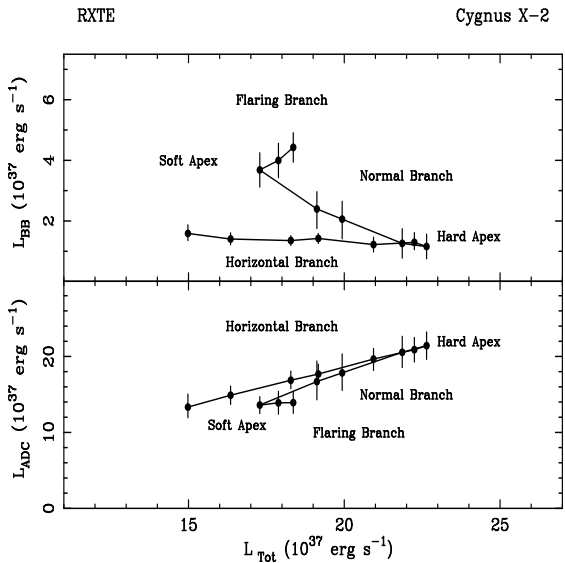}                     
\caption{Luminosities of the continuum emission components: neutron star blackbody emisssion (upper panel) 
and ADC Comptonized emission (lower panel).}
\end{center}
\end{figure}
Next in Fig. 4, we show the individual luminosities of the neutron star blackbody ($L_{\rm BB}$) and 
the Comptonized emission ($L_{\rm ADC}$) as a function of the total 1 - 30 keV luminosity, so that
neglecting the small contribution of the line: $L_{\rm Tot}$ = $L_{\rm BB}$ + $L_{\rm ADC}$. 
It can be seen that the Comptonized emission 
is the dominant component, ten times more luminous than the blackbody, and that as the source moves 
up the Z-track from the soft apex to the hard apex, this component doubles in luminosity. The X-ray 
intensity also, of course, increases by a similar factor and we suggest therefore that the mass 
accretion rate increases
in this direction contrary to the widely-held view that $\dot M$ increases monotonically round the
Z-track in the direction HB - NB - FB (Priedhorsky et al. 1986). These variations will be discussed
in more detail in the next section in terms of all 3 Cygnus\th X-2 like sources, given their
similarity of behaviour.

\section{The three sources: Cyg\th X-2, GX\th 340+0 and GX\th 5-1}

We now present the results for Cygnus\th X-2 in comparison with those we previously obtained
for GX\th 340+0 and GX\th 5-1 (Church et al. 2006; Jackson et al. 2009). Fig. 5 (upper panel) shows
the neutron star blackbody parameters. In all cases the temperature is lowest at the soft apex and
rises continuously on the NB and HB. The blackbody radius at the soft apex in all sources lies between
10 and 12 km (the dashed lines) suggesting that the whole neutron star is emitting. If this is assumed,
the results provide a value of the neutron star radius based on the three measurements 
\begin{figure}[!ht]                                                               
\begin{center}
\includegraphics[width=74mm,height=74mm,angle=270]{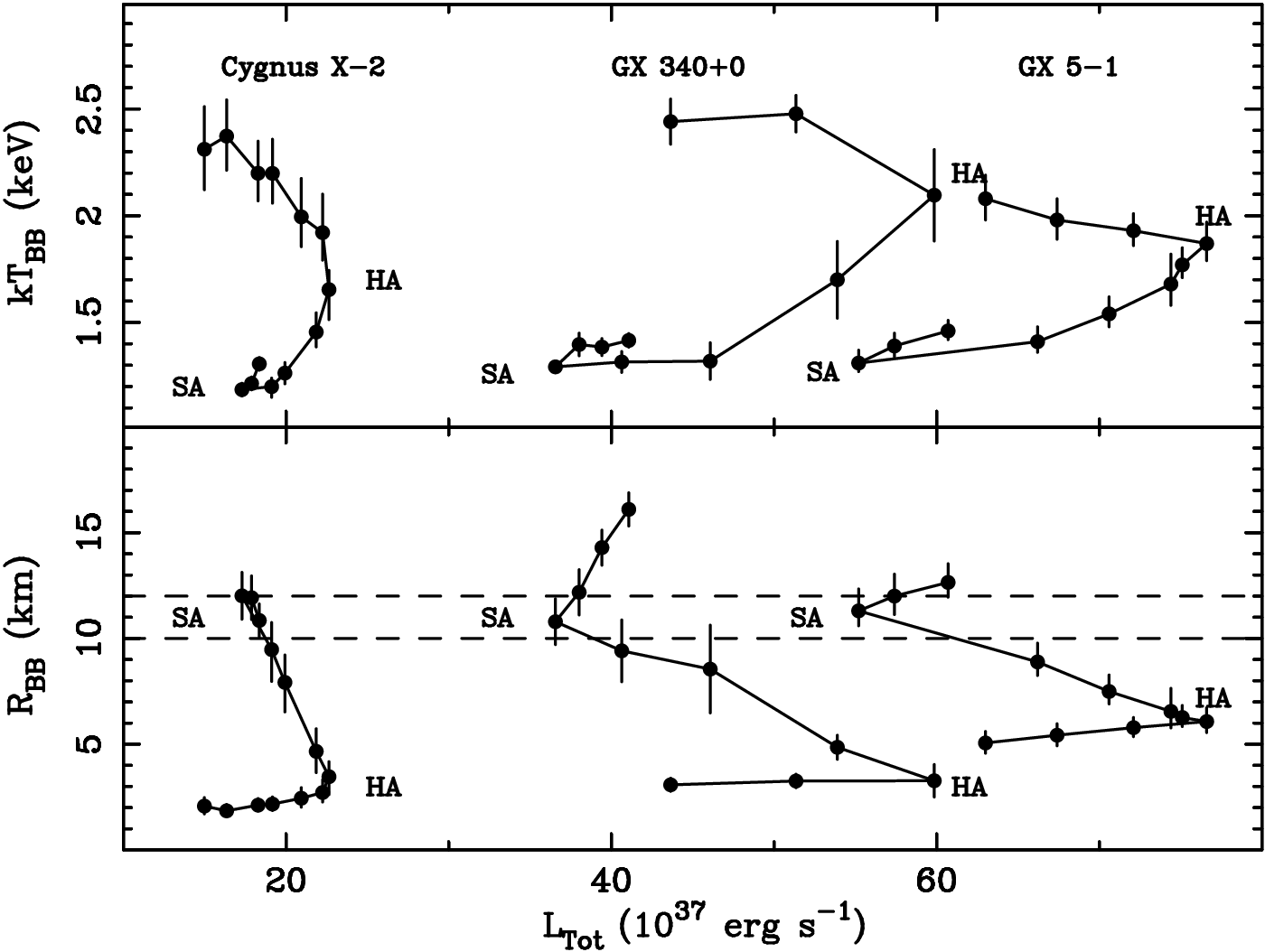}                     
\includegraphics[width=74mm,height=74mm,angle=270]{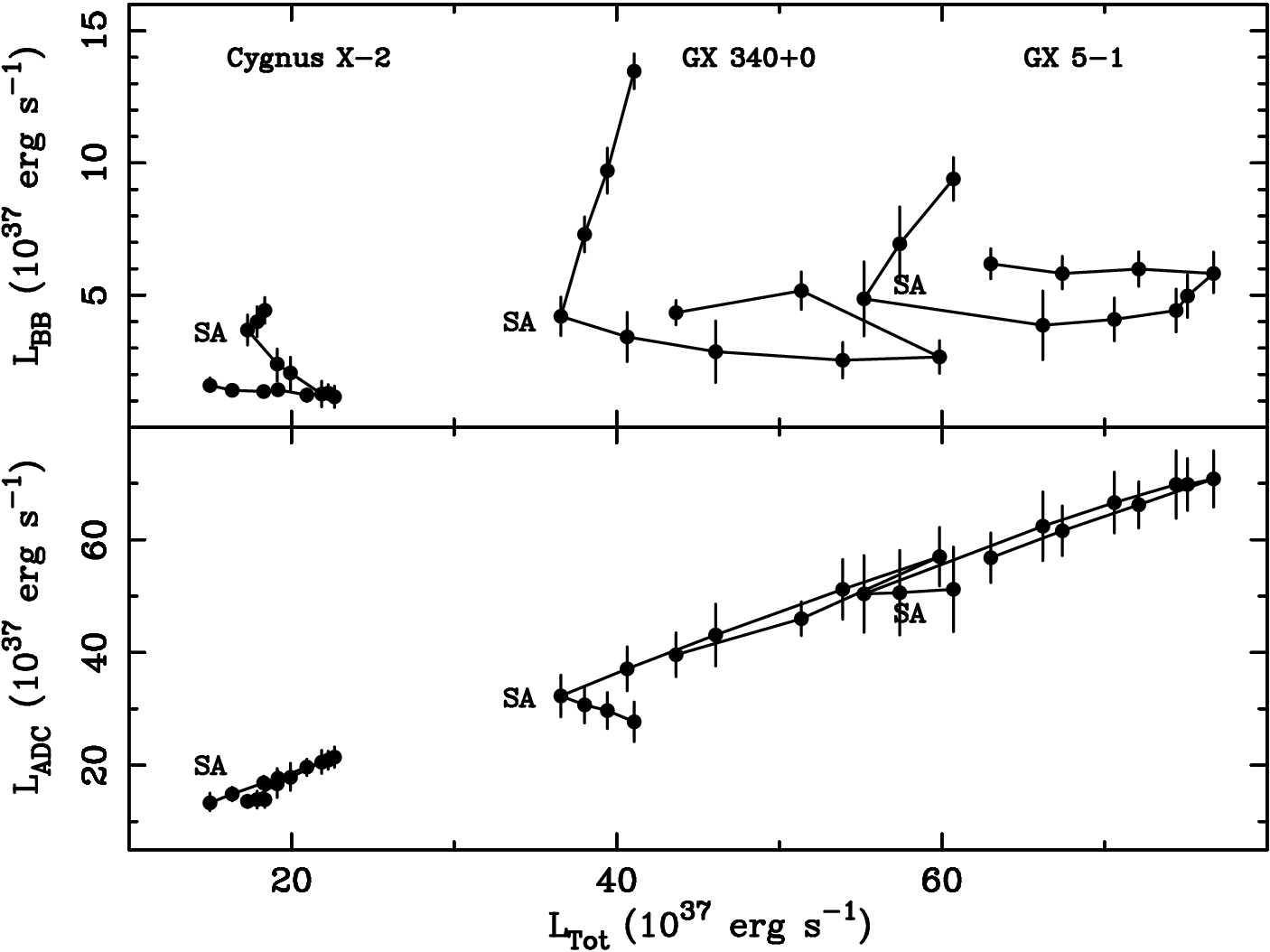}                     
\caption{Upper panel: blackbody temperature (upper panel) and radius (lower panel) as a function of the
total luminosity; lower panel: the individual luminosities of the neutron star blackbody (upper panel)
and of the Comptonized ADC emission as a function of the total luminosity (lower panel).
The soft apex (SA) and hard apex (HA) are marked.}
\end{center}
\end{figure}
of 11.4 $\pm$ 0.6 km at 90\% confidence. The flaring branch is characterized by some increase 
of temperature and a marked increase of blackbody radius.

The luminosities of the neutron star blackbody $L_{\rm BB}$ and of the Comptonized 
emission $L_{\rm ADC}$ are shown in the lower panel. In the variation of $L_{\rm ADC}$ with $L_{\rm Tot}$
the three sources lie on a single line on the NB, suggesting that the sources differ only in 
their mass accretion rate, i.e. total luminosity. We make the standard assumption of accretion physics
\begin{figure}[!ht]                                                                 
\begin{center}
\includegraphics[width=64mm,height=70mm,angle=270]{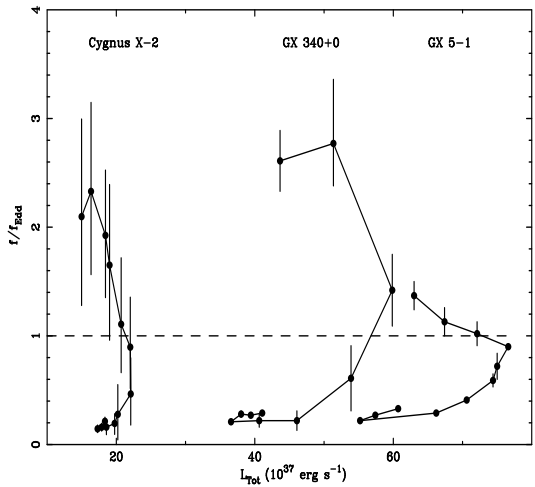}                        
\caption{Flux of the emitting part of the neutron star as a fraction of the Eddington 
flux (assumed constant): $L_{\rm Edd}/ 4\,\pi\,R^2$, where $R$ = 10 km.}
\label{}
\end{center}
\end{figure}
that the total luminosity of the source $L_{\rm Tot}$ depends on the mass accretion rate $\dot M$.
A single line is, of course expected, since $L_{\rm ADC}$
comprises a large fraction of the total luminosity. The single line does suggest that if, for example
Cygnus\th X-2 became brighter it would appear similar to GX\th 340+0 in its X-ray
properties. The increase
of $L_{\rm ADC}$ in each source between the soft apex and the hard apex strongly suggests that
the mass accretion rate $\dot M$ is increasing contrary to the standard view that $\dot M$
increases monotonically around the Z-track in the direction (HB - NB - FB). The observed increase
of neutron star blackbody temperature on the normal branch is consistent with this as we would
expect on the basis of simple accretion theory a higher temperature when more accretion flow reaches
the surface of the neutron star. On the HB the luminosity of the Comptonized 
emission falls back towards its initial value. On the NB and HB there is an increase of blackbody 
temperature, but a decrease of blackbody radius. Consequently the blackbody luminosity increases
less than expected for a temperature increase alone ($\sim$ $T^4$). 
In flaring, there is an increase of $kT_{\rm BB}$, although the flaring is weak in Cyg\th X-2.
In all three sources, $L_{\rm ADC}$ remains constant or falls
in flaring, indicating that $\dot M$ does not increase. However, the blackbody luminosity
increases substantially in the strong flaring in GX\th 340+0 and GX\th 5-1 without an
increase in $\dot M$. Thus the increase of $L_{\rm Tot}$ is due to the increase in $L_{\rm BB}$
and this must clearly be caused by an additional energy source on the neutron star, i.e. unstable nuclear burning.

\subsection{Radiation pressure}

We next view the results obtained in terms of the radiation pressure of the neutron star
which we can quantify based on the spectral fitting results. In all three sources $kT_{\rm BB}$
increases strongly on the NB and HB, by about a factor of two. Thus radiation pressure
$\sim$T$^4$ increases by nearly an order of magnitude. Moreover, as the source moves from the 
soft to the hard apex $R_{\rm BB}$ decreases, so that the emission of the neutron star becomes 
concentrated, arising from an equatorial strip, from which the emissive flux becomes high.
We consider the strength of the radiation pressure in terms of the ratio of the emissive
flux $f$ to the Eddington value $f_{\rm Edd}$ (Sect. 1). Fig. 6 shows that this ratio is initially small
at the soft apex ($\sim$0.2), i.e. sub-Eddington, but increases rapidly on the normal branch
becoming equal to unity at the hard apex, then continuing to increase on the horizontal branch
having super-Eddington values as high as $f/f_{\rm Edd}$ = 3 in GX\th 340+0. The positions 
on the Z-track where the sources are super-Eddington are exactly those positions where
radio emission is detected indicating the presence of jets, and so we propose that high radiation  
pressure plays a major role in launching the jets, by disrupting the inner disk and diverting 
accretion flow into the vertical direction. In a more detailed discussion (Church et al. 2006), 
it is shown that the reduction in blackbody radius on the NB and HB is consistent with this 
disruption. 

\subsection{Unstable nuclear burning}

In Fig. 7, we show results derived from spectral fitting for the mass accretion rate per unit area
of the neutron star $\dot m$, using $\dot M$ from the 1 - 30 keV source luminosity and dividing this 
by $4\,\pi\,R_{\rm BB}^2$. The critical value $\dot m_{\rm ST}$ at which there is a transition from 
stable nuclear burning ($\dot m$ $>$ $\dot m_{\rm ST}$) to unstable burning at smaller values of 
$\dot m$ (Bildsten 1998) is shown as a dashed line with its estimated 30\% uncertainties (dotted). 
It may be seen that there is good agreement between the values of $\dot m$ at the soft apex and $\dot m_{\rm ST}$
for the three sources suggesting that as a source descends from the hard apex towards the
soft apex, nuclear burning becomes unstable and the result is the onset of flaring. On the flaring 
branch, the constancy of $L_{\rm ADC}$ (Fig. 5) indicates that the mass accretion rate is constant.
The increase in source luminosity is due to an increase in $L_{\rm BB}$ at constant mass accretion 
rate and shows that there is an additional energy source on the neutron star not due to an increase of
accretion rate, and  nuclear burning is an obvious possibility. The above agreement with theory
indicates that this is unstable nuclear burning. This is further discussed in Sect. 6. 
During flaring there is an increase of blackbody radius (Fig. 5) from the value at the soft apex 
which we take to be the neutron star radius. These increases, e.g. to 17 km in GX\th 340+0, indicate 
an expansion of the unstable nuclear burning region beyond the surface of the neutron star similar to radius 
expansion in the burst sources.
\begin{figure}[!ht]                                                                        
\begin{center}
\includegraphics[width=65mm,height=70mm,angle=270]{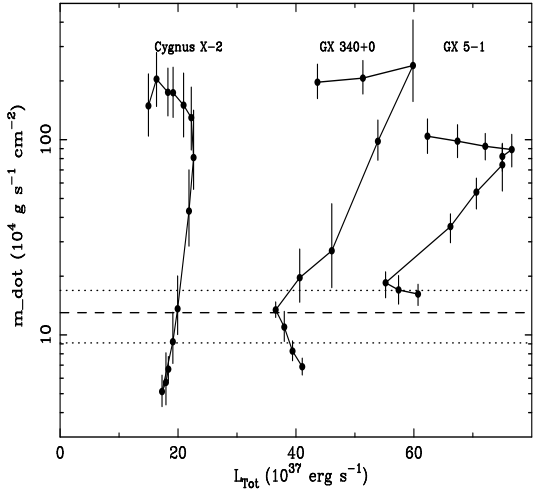}                               
\caption{Mass accretion rate per unit neutron star emitting area $\dot m$.}
\label{}
\end{center}
\end{figure}

\section{The kHz QPO}

We next present results on the kHz QPO in \hbox{Cygnus\th X-2} and GX\th 340+0
and compare these with our previous
results for GX\th 5-1 (Jackson et al. 2009). The approach adopted for GX\th 5-1 was to use 
the same selections of data made around the Z-track for both spectral analysis and timing 
analysis, thus allowing the QPO results to be viewed in terms of the spectral and physical 
evolution on the Z-track. In the case of Cygnus\th X-2, the same general approach was used, 
however, because of the relative weakness of the kHz QPO, it was necessary to increase the
size of the selection boxes (from those in Fig. 2) to allow detection of the QPO. 
Spectral fitting was again carried out for the new, larger boxes. The results of this were, 
however, entirely consistent with the results of Sect. 3, and were used in analysis of the 
QPO results. In Fig. 8 we show the new selection boxes labelled
\begin{figure}[!ht]                                                           
\begin{center}
\includegraphics[width=70mm,height=70mm,angle=270]{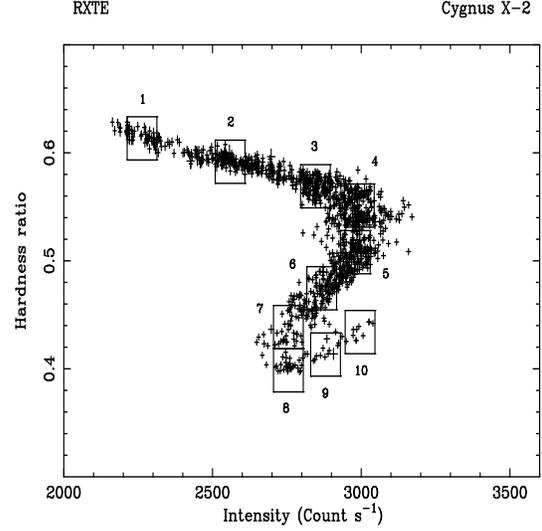}                  
\caption{Selection boxes used for fitting of the power spectra.}
\label{}
\end{center}
\end{figure}
from 1 to 10. As is well-known, kHz QPO are only detected on the HB, at the hard apex and
sometimes at the upper part of the NB, so the selection numbering begins at the end of
the Z-track where kHz QPO are expected. 

\begin{figure}[!ht]                                                           
\begin{center}
\includegraphics[width=64mm,height=64mm,angle=270]{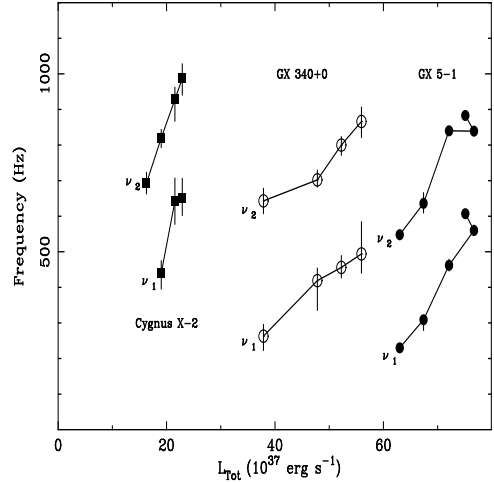}      
\caption{Kilohertz QPO frequencies as a function of source luminosity for Cygnus\th X-2,
GX\th 340+0 and GX\th 5-1.}
\label{}
\end{center}
\end{figure}

In the case of GX\th 340+0, timing analysis of the observation of 1997 September
for which we previously presented spectral fitting results (Church et al.
2006) revealed the presence of kHz QPO on the HB. However, because of a sparsity of data
in the centre of the horizontal branch, a sufficiently good dependence of kHz QPO frequency
on branch position could not be obtained. Consequently the observation of 1997 June 6 - 10 was also
analysed which was reasonably close in time to the original data. Spectral analysis as a function
of Z-track position produced spectral fitting results very similar to those presented in
Figs. 5, 6 and 7. Better coverage of the HB was available, and we used this observation
carrying out both spectral and timing analysis as a function of position on the Z-track.

In both Cyg\th X-2 and GX\th 340+0,
power spectra were extracted as described in Sect. 2 and fitted in the range 150 - 4096 Hz,
using a model consisting of two Lorentzian lines for the kHz QPO plus a power law which was able 
to fit the continuum adequately within this frequency range. KHz QPO could be detected on the 
horizontal branch only; in Cyg\th X-2 only the higher frequency kHz QPO could be detected in
box 1 (at the end of the Z-track).
Figure 9 shows the best-fit lower QPO ($\nu_1$) and upper QPO ($\nu_2$) frequencies 
as a function of the total 1 - 30 keV luminosity which was obtained by spectral fitting using the same model as in 
Sect. 4. The results for Cyg\th X-2 and GX\th 340+0 are compared with the corresponding results 
for \hbox{GX\th 5-1} (Jackson et al. 2009) also shown in Fig. 9. All three sources display an 
increase of QPO frequency with luminosity during movement along the horizontal branch
as is well-known from previous work (e.g. Jonker et al. 2000; Wijnands et al. 1998a,b). The main 
aim is, of course, to find the cause of the frequency change.

The frequency $\nu_2$ is known to correspond to the orbital frequency in the accretion disk
at positions a little way into the disk and we next view the results in these terms.
In each case we derive the radial position $r$ at which the Keplerian orbital frequency equals the
measured QPO frequency $\nu$, using the relation 
\[2\, \pi \, \nu = \sqrt{{G\, M\over r^3}}.\]
\noindent
Thus the radius $r$ decreases with increasing $L_{\rm Tot}$ for both upper and lower QPO.
We show in Fig. 10 the radial position
$r$ not as a function of luminosity, but as a function of the ratio $f/f_{\rm Edd}$
which is a measure of the strength of the radiation pressure close to the neutron star
(Sect. 4.1). The figure includes our results previously published for GX\th 5-1 (Jackson et al. 2009).
In the case of GX\th 5-1, a strong increase of the QPO radius took place when the increasing blackbody
temperature moving from the hard apex on the HB caused a marked increase in $f/f_{\rm Edd}$
at about $f$ = $f_{\rm Edd}$.
Super-Eddington values of the emissive flux suggest disruption of the inner disk
which will increase with $f/f_{\rm Edd}$ and if the upper frequency oscillation takes
place at the inner disk edge, the radius $r$ will move outwards as observed.

\begin{figure}[!ht]                                                           
\begin{center}
\includegraphics[width=64mm,height=64mm,angle=270]{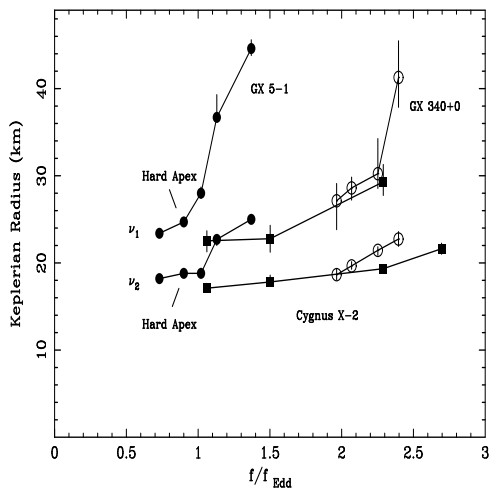}                 
\caption{Keplerian radial positions corresponding to the upper kHz QPO ($\nu_2$)
and lower kHz QPO ($\nu_1$) as a function of the strength of the neutron star radiation pressure
expressed as $f/f_{\rm Edd}$ (see text).}
\label{}
\end{center}
\end{figure}
In Cygnus\th X-2 and GX\th 340+0, the radial position $r$ also increases with $f/f_{\rm Edd}$
as in GX\th 5-1.
However, in these sources kHz QPO could only be detected on the horizontal branch so there 
are no QPO detections for $f/f_{\rm Edd}$ $<$ 1. In these sources, it appears that higher
values of $f/f_{\rm Edd}$ are needed to produce a substantial change in QPO radius, and a possible 
reason for this is discussed in Sect. 6. However, the results do suggest an explanation of 
the kHz QPO and their variation along the HB based on our spectral fitting results. In this 
explanation the disk is disrupted on the upper normal and horizontal branches as the neutron 
star emissive flux becomes super-Eddington. The higher frequency QPO with frequency $\nu_2$ 
(having the smaller radial position) is at all times an oscillation on the 
inner edge of the disk which moves outwards with increasing radiation pressure and increasing 
disk disruption, causing the change in the measured QPO frequency. The lower frequency QPO 
($\nu_1$) corresponding at the hard apex to $r$ $\sim$20 km also moves to much larger radial 
positions $\sim$30 - 40 km (Fig. 10). The mechanism producing this oscillation is not clear, 
but as its variation follows that of $\nu_2$, it seems clear that its frequency 
is derived from, i.e. depends on, $\nu_2$

\section{Discussion}

We have shown that application of the extended ADC emission model for LMXB provides good
fits to the spectra of all three Cygnus\th X-2 like Z-track sources at all positions along the Z-track. 
Moreover, the physical interpretation of the results is straightforward and suggests a plausible
explanation of the Z-track phenomenon in this group of Z-track sources. Our explanation of the 
Z-track is that the soft apex is the lowest luminosity state of the source with a low value of 
$\dot M$, with emission taking place from the whole neutron star which has its lowest temperature. 
On the normal branch, the increase of intensity and ADC luminosity suggest an increase of $\dot M$ 
leading to  heating of the neutron star and a strong increase in radiation pressure close to the neutron 
star as shown by the increase of $f/f_{\rm Edd}$ (Fig. 6), which continue on the HB.
We suggest that this has a strong effect on the inner accretion disk causing disruption 
of the disk. The horizontal force may not directly remove matter from the disk, but because 
the unperturbed height of the inner disk in LMXB at these luminosities greater than 
10$^{38}$ erg s$^{-1}$ is 20 - 50 km, the radiation pressure can also act in a direction 
close to vertical blowing away material from the upper layers of the disk. For the strongly 
super-Eddington fluxes that we measure close to the equatorial emitting zone of the
neutron star, a fraction of the mass accretion flow may be diverted vertically upwards
and be ejected from the system as massive blobs of plasma forming the jets above and below 
the disk. Thus we propose that high radiation 
\begin{table}
\begin{center}
\caption{The four r\'egimes of stable or unstable nuclear burning from Bildsten (1998),
showing the critical values of $\dot m$ that demarcate the r\'egimes.}
\begin{minipage}{80mm}
\begin{tabular}{rrrr}
\noalign{\smallskip\hrule\smallskip}
\noalign{\vskip -0.8 mm\hrule\smallskip}
1        &   2     &   3        &4\\
CNO      &Hot CNO  &He unstable &H,He   \\
unstable &stable   &in H-rich   &stable \\
         &         &environment &        \\
\noalign{\smallskip\hrule\smallskip}
\end{tabular}\\
Critical values of $\dot m$: \\
$1\times 10^3$ g cm$^{-2}$ s$^{-1}$ (r\'egimes 1 - 2)\\
$5\times 10^3$  g cm$^{-2}$ s$^{-1}$ (r\'egimes 2 - 3)\\
$1.3\times 10^5$ g cm$^{-2}$ s$^{-1}$ (r\'egimes 3 - 4)\\
\smallskip
\hrule
\smallskip
\end{minipage}
\end{center}
\end{table}
pressure is a {\it necessary} condition for 
jet formation. There are large differences
in luminosity between the Z-track sources with Cygnus\th X-2 being typically 4 times less
luminous than Sco\th X-1. Although the sources are detected in radio, in Cygnus\th X-2
jets have not so far been imaged as in Sco\th X-1 (Fomalont et al. 2001), so there is the
possibility that a second requirement for strong jet formation is a higher value of $\dot M$.

The results demonstrate that in all of the three Cygnus\th X-2 like Z-track sources, the flaring
branch consists of a strong increase of neutron star blackbody luminosity while the Comptonized
emission luminosity is constant (or decreases slightly) showing that $\dot M$ is constant.
This indicates that the FB consists of unstable nuclear burning on the neutron star.
In the theory of unstable nuclear burning (Fujimoto et al. 1981; Fushiki \& Lamb 1987; Bildsten 1998; 
Schatz et al. 1999)
the physical conditions in the atmosphere of the neutron star depend on the mass accretion
rate per unit area $\dot m$, i.e. $\dot M$ divided by the emitting area. The various r\'egimes
of stable and unstable burning are shown in Table 2. The theory applies to a wide range
of mass accretion rates, although it was developed to describe X-ray bursting at lower mass accretion rates,
and also applies to high mass accretion rates typical of the Z-track sources, and
the transition between r\'egimes 3 and 4 takes place for luminosities of about the Eddington limit.
In r\'egime 3, burning is unstable He burning in a mixed H/He environment (Bildsten 1998) but for $\dot m$
$>$ $\dot m_{\rm ST}$, H and He burning of the accretion flow become stable. The agreement in all three sources
of the measured $\dot m$ values (Fig. 7) with the critical value within its uncertainties
clearly supports flaring being unstable nuclear burning. The value of $\dot m$ at the
soft apex increases systematically with $L_{\rm Tot}$ suggesting that other, secondary effects
may take place. However, the results thus provide evidence in two ways that the FB in 
all three Cygnus\th X-2 like Z-track sources consists of unstable nuclear burning. There is also 
an expansion of the emitting region beyond the surface of the neutron star and we propose that
radius expansion may take place as it does in some X-ray bursts in lower luminosity sources.

The proposed explanation of the Z-track in the Cygnus\th X-2 like sources involves an increase
of $\dot M$ between the soft apex and hard apex, and a constant $\dot M$ in flaring. Thus
we note that this explanation departs from the often held view that the mass accretion rate
increases monotonically around the Z-track in the direction HB - NB - FB. 

We are also carrying out analysis of Sco\th X-1 like sources aimed at discovering the fundamental 
differences between the Cyg-like and Sco-like groups. Preliminary work on the source
GX\th 17+2 shows that the behaviour on the normal and horizontal branches is similar to that of the 
Cygnus\th X-2 like sources. However, in flaring there is clear evidence for an increase of mass 
accretion rate. Thus it appears that on both the FB and NB, $\dot M$ increases as we move away from
the soft apex, however with
rather different conditions on the neutron star, i.e. different values of $kT$ and $R_{\rm BB}$.  
Clearly, to fully understand the Sco\th X-1 like sources extensive work will be required; however,
it does appear that the difference between the two groups relates to the flaring branch, and not,
for example, to inclination.

Relevant to this is the decay of the transient LMXB XTE\th J1701-462, which following an outburst
in 2006 January, decreased by more than an order of magnitude in luminosity over about 500 days.
Lin et al. (2009a) showed that the source made a transition from a Z-track source of Sco\th X-1
like type to Atoll behaviour, these changes being apparent from hardness-intensity diagrams.
Towards the end of the decay, three X-ray bursts were observed and Lin et al. (2009b) claim 
that their results cast doubt on our association of the FB with unstable nuclear burning. However, 
the claim is not valid. Two of the bursts occurred in the time period labelled ``V'' by Lin et al.
in which the source showed clear Atoll behaviour, having a banana branch in hardness-intensity.
The other burst however occurred in period ``IV'' just before period V began, when flaring was weak 
and which in period V disappeared altogether. Although the luminosity was at that time 10\% of Eddington, 
the authors describe the source at the period of this burst as being on the FB of a Z-track state. On 
the basis that it may be unlikely that bursting and unstable nuclear burning in flaring take place at 
the same time it was claimed that our association of the FB with unstable burning must be doubted. The 
argument is not correct (Lin et al. 2009a,b) as our results and model for the Z-track specifically
refer to the Cygnus\th X-2 like sources. Thus the observation of the burst during a period when the 
source was described as being in a Sco\th X-1 like state (but normally described at such low luminosity 
as an upper banana state) does not cast doubt on our explanation of the Cygnus\th X-2 like sources 
in which the evidence that the FB consists of unstable burning is strong. 

%
\begin{figure}[!ht]                                                                 
\begin{center}
\includegraphics[width=52mm,height=64mm,angle=270]{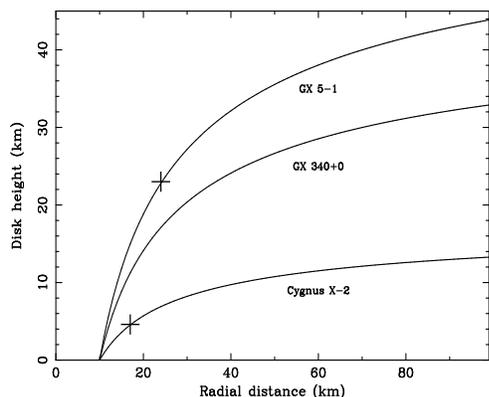}                       
\caption{Inner disk height profiles $H(r)$ for the three sources with points indicating
the position of the inner disk edge in GX\th 5-1 and Cygnus\th X-2 for $f/f_{\rm Edd}$
= 1.3 (see text).}
\label{}
\end{center}
\end{figure}
%

The kHz QPO results support the idea that high radiation pressure of the neutron star
on the upper NB and HB
disrupts the inner disk and that the upper QPO at frequency $\nu_2$ is an oscillation
actually at the inner edge of the disk. The inner radius is variable because of increasing
disruption on the horizontal branch thus causing the well-known variation of $\nu_2$ on
that branch. The fact that the radial position associated with the QPO increases when there 
is also an increase of $f/f_{\rm Edd}$ is thus suggestive. Close to the soft apex, 
$f/f_{\rm Edd}$ is much less than unity and we do not expect the disk to be disrupted at all.
Thus the inner edge of the disk will touch the neutron star, so that oscillations of the edge
will probably not be possible. Thus this model for the upper frequency kHz QPO predicts that
kHz QPO will not be detected on the lower normal branch, in agreement with observation.

In GX\th 5-1, the most luminous source, substantial change in the radial position of
both upper and lower QPO takes place at $f/f_{\rm Edd}$ $\sim$1, but in GX\th 340+0 and
Cygnus\th X-2, larger values of $f/f_{\rm Edd}$ are needed which may appear surprising.
However, the inner disks of the three sources have very different vertical heights
and height profiles $H(r)$ and these are shown in Fig. 11 based on the standard theory
of the thick disk supported vertically by its own radiation pressure 
(Frank et al. 2002). For each source, $H(r)$ is plotted to correspond to the luminosity
at the hard apex of the Z-track where we expect there to be disruption of the inner disk.
At radial positions $r$ $<$ 50 km, appropriate to the kHz QPO,
$H(r)$ is rising and eventually reaches the equilibrium height of the radiatively supported 
disk. Also marked on the figure are radial positions corresponding to a fixed value of $f/f_{\rm Edd}$
= 1.3. Fig. 10 shows that for this value, the upper kHz QPO has an equivalent orbital
radius of 24 km GX\th 5-1 and 17 km in Cyg\th X-2 (there is no value of $f/f_{\rm Edd}$
that includes all three 
sources). Thus the two points plotted in Fig. 10 indicate the radial positions where the QPOs 
occur, which we assume are the inner edges of the disk and so indicate the radial and vertical
extent of disk disruption. This shows that in the brighter source GX\th 5-1 the disruption
for a given level of radiation pressure is greater. A theoretical treatment of disruption
is complex and beyond the scope of this work; however, it does not appear unreasonable
that in a source with greater disk height (GX\th 5-1), the disruption is greater.

\thanks{
This work was supported in part by the Polish Ministry of Science and Higher Education grant 3946/B/H03/2008/34.}

\end{document}